\documentclass[aps,prl,showpacs,twocolumn]{revtex4}
\usepackage{graphicx}
\usepackage{amssymb}
\usepackage{amsmath}
\usepackage{bm}
\begin{document}

\title[]{Collisional Dynamics of Half-Quantum Vortices in a Spinor Bose-Einstein Condensate}

\author{Sang Won Seo$^{1}$}
\author{Woo Jin Kwon$^{1,2}$}
\author{Seji Kang$^{1,2}$}
\author{Y. Shin$^{1,2}$}
\email{yishin@snu.ac.kr}

\affiliation{
$^1$Department of Physics and Astronomy, and Institute of Applied Physics, Seoul National University, Seoul 08826, Korea\\
$^2$Center for Correlated Electron Systems, Institute for Basic Science, Seoul 08826, Korea
}

%\date{\today}

\begin{abstract}
We present an experimental study on the interaction and dynamics of half-quantum vortices (HQVs) in an antiferromagnetic spinor Bose-Einstein condensate. By exploiting the orbit motion of a vortex dipole in a trapped condensate, we perform a collision experiment of two HQV pairs, and observe that the scattering motions of the HQVs is consistent with the short-range vortex interaction that arises from nonsingular magnetized vortex cores. We also investigate the relaxation dynamics of turbulent condensates containing many HQVs, and demonstrate that spin wave excitations are generated by the collisional motions of the HQVs. The short-range vortex interaction and the HQV-magnon coupling represent two characteristics of the HQV dynamics in the spinor superfluid.
\end{abstract}

\maketitle

When a superfluid has an internal spin degree of freedom, quantum vortices (QVs) with fractional values of the quantum circulation, $h/m$, may exist in the superfluid due to the complex topology of the order parameter manifold~\cite{RJD, Volovik}, where $h$ is Planck's constant and $m$ is particle mass. Recently, QVs with $h/2m$, known as half-quantum vortices (HQVs), were reported with many spinor superfluid systems; $\pi$ rotation of light polarization was shown around defects in exciton-polariton condensates~\cite{Lagoudakis,Manni}, half-height magnetization steps were detected in a ring-shaped spin-triplet superconductor of Sr$_2$RuO$_4$~\cite{Jang}, and spontaneous dissociation of a singly charged vortex into a pair of HQVs with ferromagnetic cores was observed in an antiferromagnetic spinor Bose-Einstein condensate (BEC)~\cite{Seo}. More recently, HQVs were identified in NMR measurements of a rotating superfluid $^3$He in the polar phase~\cite{Autti}.

Given the stable existence of HQVs  in the spinor superfluids, the next immediate questions concern their dynamic properties~\cite{HQVcoresize,Eto,Kaneda,Kasamatsu,Tsubota}. Since a HQV intrinsically involves a spin texture, the interplay  between mass and spin flows will play an important role in the HQV dynamics. In particular, a HQV has a nonsingular core structure where the core region is occupied by a nonrotating spin component, thus, having a continuous vorticity distribution. This is in stark contrast to the conventional QV which has a density-vanishing core with phase singularity. Therefore, the HQV dynamics would be qualitatively different from that of QVs in a scalar superfluid. It has been anticipated that HQVs can merge into and come out of topological solitons such as Skrymions and merons~\cite{Volovik2,Choi}, and spin monopoles~\cite{Ruostekoski,Ray,Tiurev}.

In this Letter, we present an experimental study on the dynamics of HQVs in an antiferromagnetic spinor BEC. By means of a vortex-dipole generation technique, we perform a collision experiment of two HQV pairs in a highly oblate BEC, and show that two HQVs with opposite core magnetizations have a short-range interaction. We also investigate the relaxation dynamics of turbulent BECs containing many HQVs, and observe that spin wave excitations are generated by the collisional motions of HQVs, revealing the dissipative mechanism in the HQV dynamics. The short-range interaction of HQVs and the coupling between HQVs and magnons represent the prominent features of the HQV dynamics. Our findings manifest  the dynamic interplay of the mass and spin sectors in the spinor superfluid system.

We study a BEC of $^{23}$Na atoms in the $F=1$ hyperfine spin state. The spin interaction in the spinor condensate is antiferromagnetic, whose energy is given as $E_\mathrm{S}=\frac{c_2 n}{2}\langle \mathbf{F}\rangle^2$ with $c_2>0$, where $n$ is the atomic density and $\mathbf{F}=(F_x,F_y,F_z)$ is the single-particle spin operator. The ground spin state is polar with $\langle \mathbf{F}\rangle=0$ that can be parametrized with a unit vector $\vec{d}$, such that the system is in the $m_F=0$ spin state for the quantization axis along $\vec{d}$~\cite{Ho,Machida}. The order parameter of the condensate is expressed as
\begin{align}
\Psi_{\mathrm{AF}}=\left(
\begin{matrix}
  \psi_{+1}\\
  \psi_{0}\\
  \psi_{-1}
 \end{matrix}\right)=\sqrt{n}e^{i\theta}\left(
\begin{matrix}
  \frac{-d_x+i d_y}{\sqrt{2}}\\
  d_z\\
  \frac{d_x + i d_y}{\sqrt{2}}
\end{matrix}\right),
\end{align}
where $\psi_l$ is the $m_z=l$ spin component along the $z$ direction ($l=0,\pm1$) and $\theta$ is the superfluid phase. The order parameter is invariant under $\theta\rightarrow\theta+\pi$ and $\vec{d}\rightarrow -\vec{d}$. This $\mathrm{Z}_2$ symmetry allows HQV formation consisting of $\pi$ phase winding together with spin flipping around the vortex core~\cite{Zhou,Kawaguchi_review, Stamper-kurn_review}.

In the presence of an external magnetic field, e.g., along the $z$ direction, the quadratic Zeeman shift, $q$, introduces a uniaxial spin anisotropy to the system with the energy of $E_\mathrm{Z}=q\langle F_z^2\rangle=q(1-d_z^2)$. The ground state has a spin orientation $\vec{d}\parallel \hat{z}$ for $q>0$ and $\vec{d}\perp\hat{z}$ for $q<0$. We refer to these two phases as the easy-axis polar (EAP) and easy-plane polar (EPP) phases, respectively. In the EPP phase, the spin can rotate in the $xy$ plane, admitting formation of HQVs having two topological charges, $q_n$ and $q_s$ ($|q_n|=|q_s|=\frac{1}{2}$), which are the winding numbers of $\theta$ and $\vec{d}$, respectively. The EPP phase is an equal mixture of the $m_z=+1$ and $-1$ spin components, and HQV can be understood as a quantum vortex in one spin component whose core is occupied by the other nonrotating spin component and, thus, is magnetized. The characteristic size of the ferromagnetic core is given by the spin healing length $\xi_s=\hbar/\sqrt{2m c_2 n}$~\cite{HQVcoresize,Eto}.

\begin{figure}
\centering
\includegraphics[width=8.0cm]{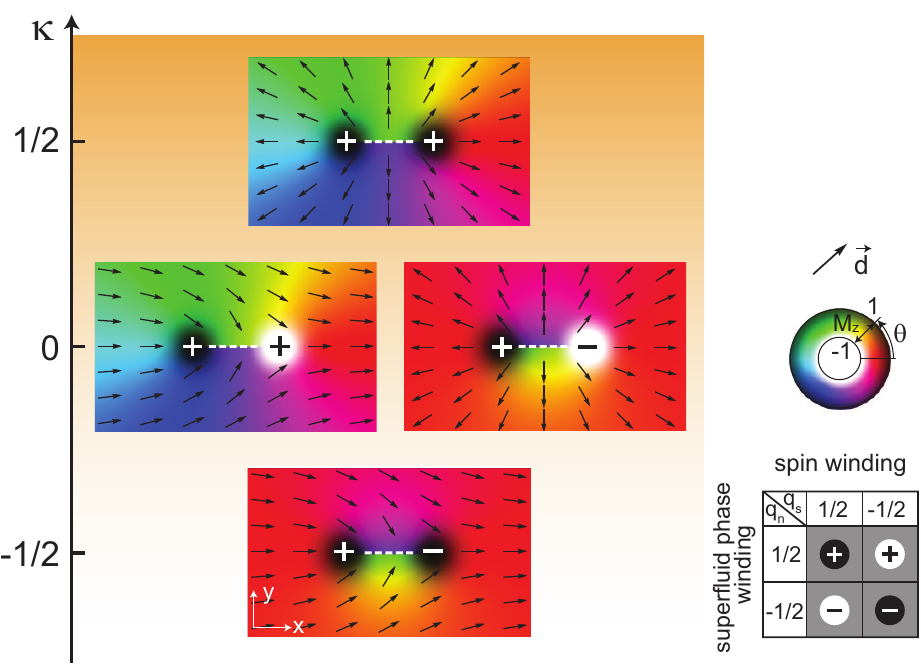}
\caption{HQV pairs in an antiferromagnetic spinor BEC in the easy-plane polar (EPP) phase. The diagrams show the spatial structures of four HQV pairs in terms of the superfluid phase $\theta$ (color) and the spin orientation $\vec{d}$ (arrow). The magnetization $M_z$ in the core region is indicated by brightness. The topological charges of the left and right HQVs are $(q_{n1},q_{s1})=(\frac{1}{2},\frac{1}{2})$ and $(|q_{n2}|,|q_{s2}|)=(\frac{1}{2},\frac{1}{2})$, respectively, where $q_n$ and $q_s$ correspond to the winding numbers of $\theta$ and $\vec{d}$, respectively. $\kappa=q_{n1} q_{n2}+q_{s1} q_{s2}$. The dashed line between the two HQVs denotes a sudden change of $\theta\rightarrow\theta+\pi$ and $\vec{d}\rightarrow -\vec{d}$. The superfluid order parameter is continuous across the dashed line due to the $\mathrm{Z}_2$ symmetry (see text). The HQV pair with $\kappa=-1/2$ is a HQV dipole with zero net charges.}
\label{fig1}
\end{figure}

\begin{figure}
\includegraphics[width=8.5cm]{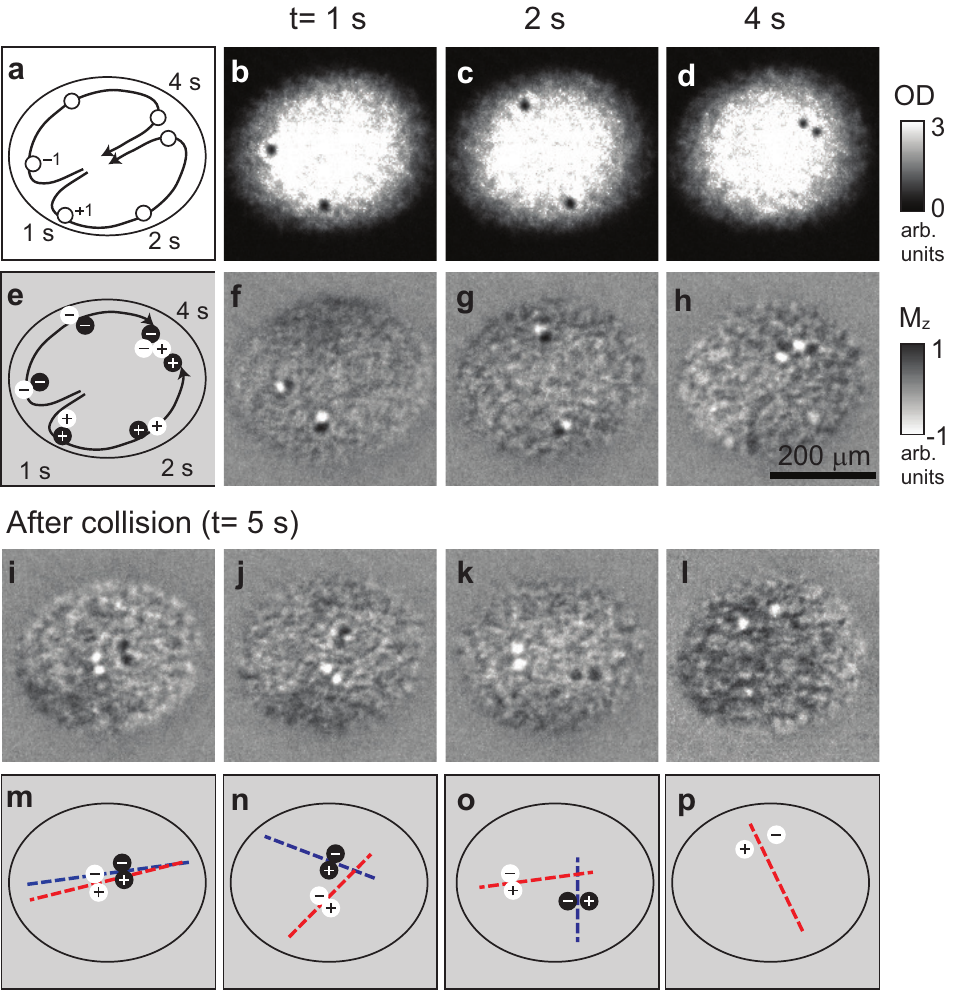}
\caption{Collision of HQV pairs. (a) A vortex dipole consisting of two singly charged vortices with opposite circulations is generated in the center region of the condensate~\cite{Supple}. The two vortices show an orbit motion depicted by the solid lines. (b)-(d) Optical density (OD) images of BECs in the easy-axis polar (EAP) phase, taken at (b) $t=1$~s, (c) 2~s, and (d) 4~s after generating a vortex dipole. (e) When the BEC is transmuted to the EPP phase after the vortex generation, each singly charged vortex is split into a pair of HQVs with opposite core magnetizations, and the HQV pair moves along the trajectory of its original singly charged vortex. (f)-(h) Magnetization images of BECs in the EPP phase at (f) $t=1$~s, (g) 2~s, (h) 4~s, and (i)-(l) 5~s. At $t\approx4$~s, two HQV pairs collide in the upper right region of the BEC and scatter into two HQV dipoles. (m)-(p) Descriptions of the vortex states in (i)-(l). The dashed lines indicate the propagation lines of the HQV dipoles. In (l), only one HQV dipole is identified. All images were taken after a free expansion of the samples for 24~ms.}
\label{fig2}
\end{figure}

The interaction between vortices arises mainly from the interference of the velocity fields generated by the vortices. Much like for QVs in a scalar superfluid, the HQV-HQV interaction energy shows a logarithmic form of $\kappa \ln R$  for a large distance $R\gg \xi_s$, where $\kappa=q_{n1}q_{n2}+q_{s1}q_{s2}$, parametrizing the overall interference effect in the mass and spin currents, and ($q_{ni}$, $q_{si}$) are the charges of the $i$th HQV ($i=1,2$). When $R$ is comparable to the core size $\xi_s$, the core structure is modified in the proximity of the other vortex, consequently affecting the HQV interaction. Particularly, in the case of $\kappa=0$ where the two HQVs have different core magnetizations (Fig.~1), the core deformation is the main mechanism governing the HQV interaction. Theoretical studies predicted that the HQV interaction for $\kappa=0$ is repulsive and short-ranged with an asymptotic form of $(\ln R) /R^2$~\cite{HQVcoresize,Eto}.

Our experiment starts by preparing a nearly pure BEC of $^{23}$Na atoms in the $|F=1,m_F=0\rangle$ hyperfine spin state in an oblate optical dipole trap~\cite{Seo,Supple}. The condensate typically contains $N_a=5.6\times 10^6$ atoms with Thomas-Fermi radii $(R_x,R_y,R_z)\approx(203, 163, 1.8)~\mu$m. For peak atom density, the density and spin healing lengths are $\xi_n\approx 0.5~\mu$m and  $\xi_s\approx4.1~\mu$m, respectively. Since $\xi_s>R_z$, the spin dynamics in the oblate condensate is effectively two dimensional. The external magnetic field is $B_z=30$~mG along the $z$ direction, giving $q/h=0.24$~Hz. The value of $q$ can be tuned to be negative by using a microwave dressing technique~\cite{Gerbier,Quenching,Zhao}. HQVs are detected as ferromagnetic point defects in magnetization imaging~\cite{Seo,Supple,Higbie}, or with density-depleted cores in absorption imaging after Stern-Gerlach spin separation.

\begin{figure*}
\includegraphics[width=15.0cm]{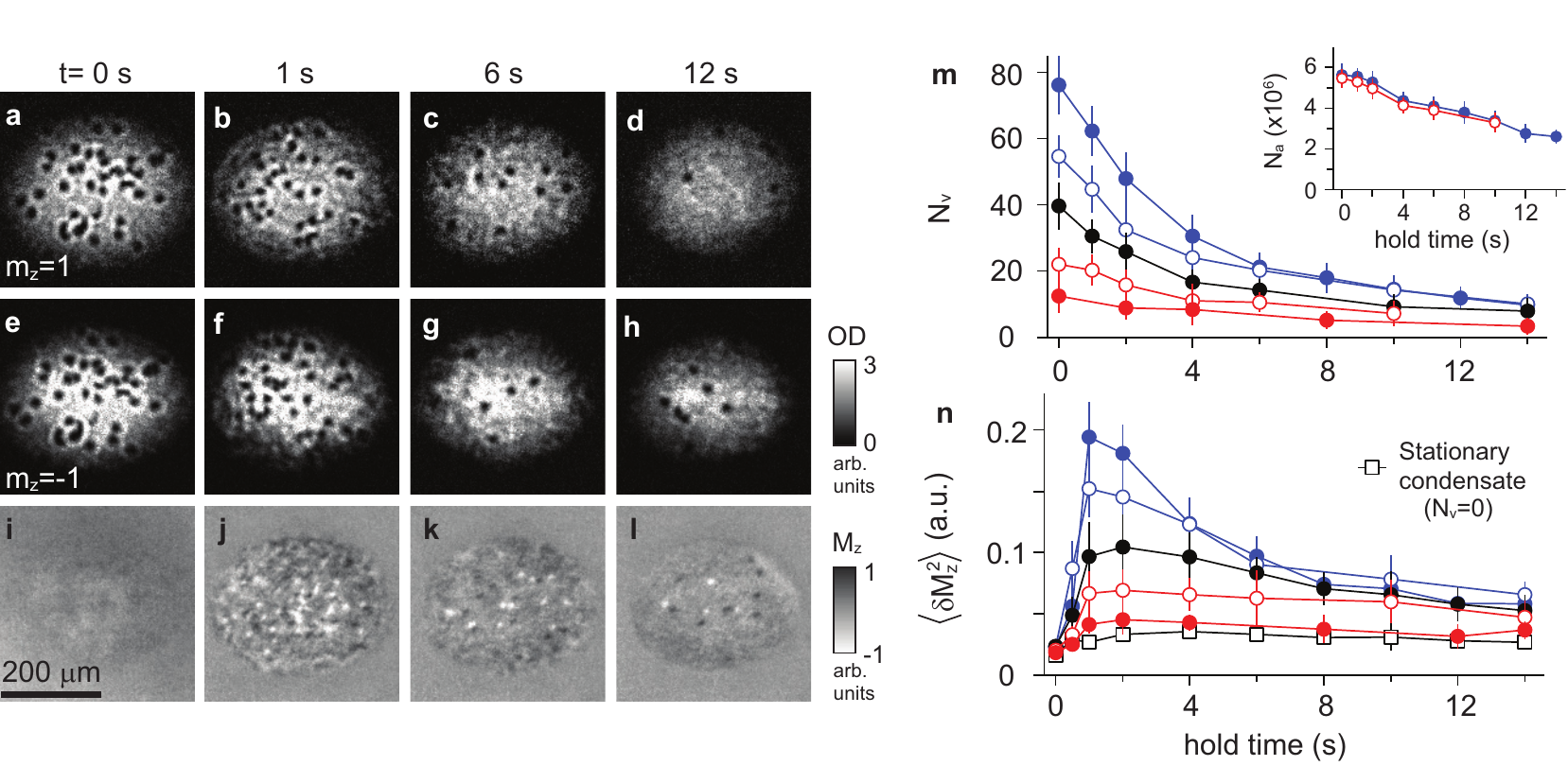}
\caption{Relaxation of quantum turbulence with HQVs. OD images of  (a)-(d) the $m_z=1$  and (e)-(h) $m_z=-1$ spin components of turbulent BECs at various hold times, $t$, in the EPP phase. The two spin components were imaged simultaneously with Stern-Gerlach spin separation and 24 ms time of flight~\cite{Supple}. The vortex positions in the two spin components coincide initially and become uncorrelated in the subsequent evolution, indicating the development of HQVs. The cloud shapes of the two spin components are slightly mismatched due to the inhomogeneity of the spin-separating field gradient. (i)-(l) {\it In situ} magnetization images of the BECs for the corresponding hold times. Spin wave excitations are rapidly generated as HQVs develop. Temporal evolutions of (m) the HQV number, $N_v$, and (n) the variance of spin fluctuations, $\langle \delta M_z^2\rangle$, for various sample conditions.  $N_v$ was measured from the OD images of the two spin components and $\langle \delta M_z^2\rangle$ was determined from  the central $150~\mu$m$\times 150~\mu$m region of  the {\it in situ} magnetization image. The data points were obtained by averaging ten measurements of the same experiment, and the error bars indicate the standard deviation of the measurements. The inset in (m) shows the atom number evolution of the BEC for initial vortex number $N_v\approx 80$  (solid) and 20 (open). Over 12~s evolution, the condensate fraction slightly decreased from $>90\%$ to $\approx 85\%$.}
\label{fig3}
\end{figure*}

We perform a collision experiment of HQVs by exploiting the orbit motion of a vortex dipole in a trapped condensate [Fig.~2(a)]~\cite{Neely,Freilich}. First, we generate a vortex dipole of two singly charged vortices with opposite circulations by sweeping the center region of the condensate with a penetrable repulsive laser beam ~\cite{Supple,Kwon1}. The vortex dipole linearly propagates, and near the condensate boundary, the two vortices separate and circulate in opposite directions along the condensate perimeter [Figs.~2(b) and 2(c)]. They eventually combine at the other side of the condensate [Fig.~2(d)] and traverse the condensate again as a small dipole. This orbit motion can be explained with the velocity field of the vortices and the boundary condition of the trapped condensate~\cite{Neely,Freilich}. In our sample, the orbit motion period was about $4.5$~s.

For a HQV collision experiment, we transmute the condensate from the EAP phase into the EPP phase immediately after the vortex generation. It is achieved by applying a $\pi/2$ rf pulse to rotate $\vec{d}$ from $\hat{z}$ to the $xy$ plane and, subsequently changing $q/h$ to $-10$~Hz. In the EPP phase, a singly charged vortex with $(q_n,q_s)=(\pm 1,0)$ is unstable and dissociated into two HQVs with $(q_n,q_s)=(\pm\frac{1}{2},\frac{1}{2})$ and $(\pm\frac{1}{2},-\frac{1}{2})$~\cite{Seo}. We observe that the splitting process is completed within 1.5~s and the two HQV pairs move along the trajectories of their original singly charged vortices, maintaining their small pair separation  [Figs.~2(f) and 2(g)].  Note that, thanks to the vortex orbit motion and the core magnetization, we can unambiguously specify the charges $(q_n,q_s)$ of each HQV [Fig.~2(e)]. The pair separation was about $4\xi_s$ and its direction was random in each realization of the experiment~\cite{Seo}. It was predicted that a HQV pair created from a singly charged vortex undergoes pair rotation due to its repulsive short-range interaction~\cite{Kasamatsu,Kaneda}.

The two HQV pairs collide in the upper right region of the condensate [Fig.~2(h)] and scatter into two HQV {\it dipoles} with zero net charges: one is a $\{(\frac{1}{2},\frac{1}{2}), (-\frac{1}{2},-\frac{1}{2})\}$ dipole with spin-up core magnetization and the other is a $\{(\frac{1}{2},-\frac{1}{2}), (-\frac{1}{2},+\frac{1}{2})\}$ dipole with spin-down core magnetization [Figs.~2(i)-2(k)]. We see that the propagation lines of the two HQV dipoles cross in the region where the four HQVs gathered [Figs.~2(m)-2(o)], recalling the HQV collision event. The observed splitting, orbiting, and scattering motions of the HQVs corroborate the existence of the short-range interaction between HQVs with different core magnetizations.

Although in as few as 4 out of about 130 runs, we made an interesting observation where only one HQV dipole remains after collision, as shown in Fig.~2(l). This implies that the other HQV dipole was annihilated during the collision process. The pair annihilation requires energy dissipation, which possibly happens via dynamic coupling of HQVs to other excitations in the system, such as phonons and magnons. In the EPP phase, there are two modes of spin excitation: the gapless axial magnons due to the broken spin rotation symmetry in the $xy$ plane, and the gapped transverse magnons associated with the $m_z=0$ spin component~\cite{Kawaguchi_review}. The ferromagnetic cores of a HQV dipole would be mainly released as axial magnons in the annihilation process. In recent numerical simulations of the HQV dipole dynamics based on 2D Gross-Pitaevskii equations, a peculiar tendency of pair annihilation was observed for short intervortex separation~\cite{Kasamatsu}.

To investigate the dissipative nature of the HQV dynamics, we examine the relaxation of a turbulent BEC containing many HQVs (Fig.~3). We prepare a condensate with a spatially disordered vortex distribution in the EAP phase~\cite{Supple,Kwon2} and transmute it into the EPP phase. Large spin wave excitations are rapidly generated as the singly charged vortices are dissociated and the resultant HQVs are spatially scrambled [Figs.~3(i) and 3(j)]. Then, the turbulent condensate gradually relaxes to a stationary state, decreasing the HQV number, $N_v$. We characterize the relaxation dynamics with the temporal evolutions of $N_v$ and the variance of spatial spin fluctuations, $\langle \delta M_z^2 \rangle$ [Figs.~3(m) and 3(n)]~\cite{Supple}. $\langle \delta M_z^2 \rangle$ reflects the axial magnon population in the condensate~\cite{Seo,Symes}. We observed no $m_z=0$ spin component in the relaxation dynamics, excluding the involvement of the gapped transverse magnons.

In Figure 4, we display the relaxation trajectories of the turbulent condensate in the plane of the dimensionless HQV density, $n_v\xi_s^2$, and $\langle \delta M_z^2 \rangle$, where $n_v=N_v/(\pi R_x R_y)$. Interestingly, we see that the turbulent condensate follows a certain universal relaxation curve after it reaches its maximum value of $\langle \delta M_z^2 \rangle$, regardless of the initial vortex number. One plausible explanation is that the magnon population is dynamically governed by the rate of magnon generation from the collisional motions of the HQVs. Along the universal relaxation curve,  the dependence of $\langle \delta M_z^2 \rangle$ on $n_v\xi_s^2$ appears faster than linear, which seems to support the explanation because the occurrence probability of HQV collision must increase faster than linearly with the vortex density.

Finally, we analyze the decay curve of $N_v$ in the turbulent condensate (Fig.~4 inset). Nonexponential decay behavior is observed, where the decay rate is reduced from 0.2~s$^{-1}$ to 0.1~s$^{-1}$ as $N_v$ decreases from 60 to lower than 10.  For comparison, we measured the decay curve of the singly charged vortex number with the same turbulent condensates in the EAP phase, and found that it decays exponentially over the entire experimental range with a decay rate of about 0.03~s$^{-1}$~\cite{Supple}. The fast and nonexponential decay behavior in the EPP phase highlights the role of the HQV-magnon coupling as a dissipative mechanism in the spinor superfluid system.

\begin{figure}
\includegraphics[width=8.0cm]{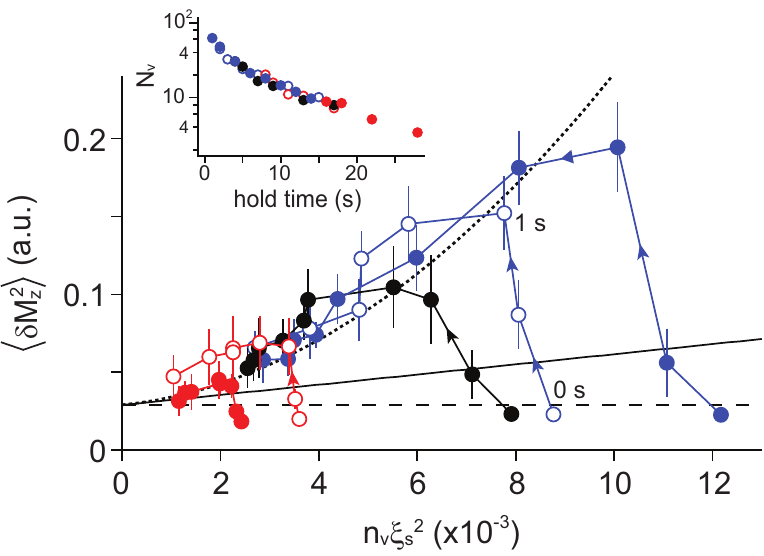}
\caption{Relaxation trajectories of turbulent BECs in the plane of the dimensionless HQV density, $n_v \xi_s^2$, and $\langle \delta M_z^2\rangle$. The arrows represent the evolution direction of the trajectories. The dashed line indicates the thermal equilibrium value of $\langle \delta M_z^2\rangle$ for the stationary condensate, which are obtained by averaging the data points marked by open squares in Fig.~3(n) for long hold times $t\geq 6$~s. The solid line shows the estimated contribution of magnetized HQV cores, $\langle \delta M_z^2\rangle_{\text{HQV}}$~\cite{Supple}, with the thermal equilibrium offset. The dotted line is a guide for the eyes. The variation of the atom number of the condensate was taken into account in the calculation of $n_v \xi_s^2$ [Fig.~3(m), inset]. The inset shows the universal decay curve of the vortex number, constructed by joining the data in Fig.~3(m)  for $t\geq 1$~s with adjusted hold time offsets.}
\label{fig4}
\end{figure}

In conclusion, we investigated the dynamics of HQVs in the antiferromagnetic spinor Bose-Einstein condensate and demonstrated the existence of the short-range interaction of HQVs and the dynamic coupling between HQVs and magnons. It would be interesting to extend this work to the quantum critical point of $q=0$, where the system recovers the full spin rotation symmetry of $S^2$ and the spin winding number, $q_s$, is not topologically defined. Furthermore, in a 2D regime, no spin ordering would form at finite temperature~\cite{Mermin66,Nelson1977}, and it was anticipated that the so-called paired superfluid state might emerge near the critical point~\cite{Mukerjee,Pietila,Phase_Diagram}.

This work was supported by the National Research Foundation of Korea (Grants No. 2011-0017527 and No. 2013-H1A8A1003984) and IBS-R009-D1.

\newpage
\section{Supplemental Material}

\noindent {\bf Sample preparation.} A cold thermal cloud of $^{23}$Na atoms in the $|F=1,m_F=-1\rangle$ hyperfine spin state was generated in an optically plugged magnetic quadrupole trap~\cite{Heo} and transferred to an optical dipole trap. A BEC of about $5.6\times 10^6$ atoms was produced by evaporation cooling and the thermal fraction of the sample was less than 10\%. The final trapping frequencies of the optical dipole trap were $(\omega_x, \omega_y,\omega_z)=2\pi\times(4.2, 5.3, 480)$~Hz. The spin state of the condensate was changed from $|m_z=-1\rangle$ to $|m_z=0\rangle$ via a Landau-Zener sweep of the radio frequency field at an external magnetic field $B_z=2.5$~G. Then, the magnetic field was ramped down to $B_z=30$~mG. The background field gradient was controlled to be less than $0.1$~mG/cm. The microwave field for tuning the quadratic Zeeman shift, $q$, was linearly polarized along the $y$ direction and its frequency was detuned by $-300$~kHz with respect to the $|F=1,m_F=0\rangle \rightarrow |F=2,m_F=0\rangle$ transition. Under the microwave dressing, the quadratic Zeeman field was estimated to be $q/h= -10$~Hz from the measurements of the Rabi frequency of the microwave field~\cite{Zhao}.

\noindent{\bf Vortex state preparation and detection.} Vortices were generated by sweeping the condensate with a repulsive Gaussian laser beam. The $1/e^2$ width of the laser beam was $\approx9~\mu$m. In the case of the single vortex-dipole generation, we set the potential height of the optical obstacle to be $V/\mu\approx$0.8, where $\mu$ is the chemical potential of the condensate, and we linearly swept a central region of the trapped condensate for 24~$\mu$m in 21~ms. The generation efficiency of a single vortex dipole was about $80\%$~\cite{Kwon1}. For turbulence generation, we stirred the condensate with the same laser beam of $V/\mu\approx 2$ in a sinusoidal manner for 200~ms~\cite{Kwon2}. The vortex number was controlled by varying the stirring amplitude and frequency of the laser beam. In this case, the condensate was in the $|m_z=-1\rangle$ state and confined in a tighter trap with trapping frequencies of $(\omega_{x,y},\omega_z)=2\pi\times(15, 400)$~Hz, which was formed by superimposing a magnetic quadrupole field to the optical dipole trap. Using the tight trap was beneficial for preparing a spatially disordered vortex distribution. After stirring, we adiabatically turned off the magnetic quadrupole field within 2 s and applied the Landau-Zener sweep of the radio frequency field to transfer the spin state to the $|m_z=0\rangle$ state.

The vortices were detected by taking time-of-flight absorption imaging with Stern-Gerlach spin separation, or {\it in-situ} or time-of-flight magnetization imaging. For the spin separation, after suddenly turning off the optical trap, we adiabatically rotated the external magnetic field to the $x$ direction and applied a field gradient along the $x$ direction for 7 ms to spatially separates the spin components. Vortices appear as density-depleted holes in the imaging. In the case of HQVs, the positions of the holes in the image of one spin component are different from those in the image of the other spin component. In the magnetization imaging, HQVs are detected as ferromagnetic point defects due to their magnetized vortex cores. Our spatial imaging resolution was about $4~\mu$m, compatible to the spin healing length $\xi_s\approx 4.1~\mu$m, and HQVs could be observed in {\it in-situ} magnetization imaging~\cite{Seo}.

\begin{figure}[b]
\includegraphics[width=6.5cm]{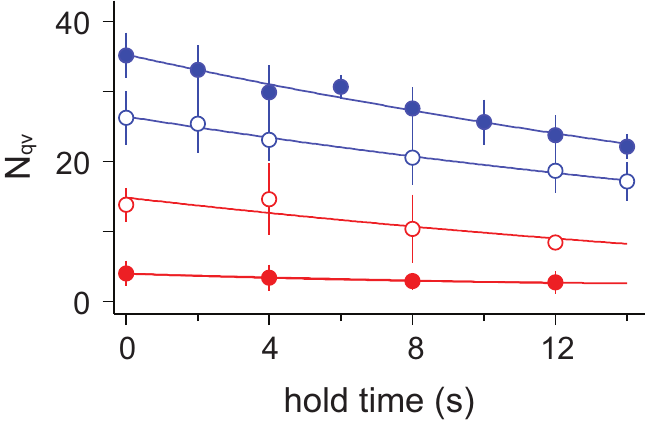}
\caption{{\bf Decay curve of the vortex number in the EAP phase}. Turbulent condensates were prepared in the same manner as in the main experiment and the decay curve of the singly-charged vortex number, $N_{qv}$, was measured in the EAP phase, i.e., without transmuting the condensates to the EPP phase. In the measurement, we set the external magnetic field to be $B_z\approx500$~mG in order to suppress spin dynamics in the condensate. The solid lines are the exponential fits to the data. The exponential decay rate was consistently obtained to be 0.034(4)~s$^{-1}$ for various initial vortex numbers.}
\label{FigS1}
\end{figure}

\noindent{\bf Magnetization imaging.} The magnetization distribution $M_z(x,y)$ of the condensate was measured with spin-dependent phase-contrast imaging~\cite{Seo}. The probe beam frequency was set at $-20$~MHz, red-detuned from the $3S_{1/2}|F=1\rangle \rightarrow 3P_{3/2}|F^{\prime}=2\rangle$ transition and the probe beam polarization was $\sigma^{-}$. In the phase-contrast imaging, the signs of the optical signals of the $m_z=1$ and $m_z=-1$ spin components are opposite and the net gain in the optical signal is proportional to the condensate magnetization, $M_z=\tilde{n}\langle F_z\rangle$, where $\tilde{n}=\int{n}dz$ is the atomic column density. The magnetization was not precisely calibrated and normalized with the peak magnetization magnitude, $|M_{z0}|$, at the HQV core.

\noindent{\bf Analysis of spin fluctuations.} The variance in spin fluctuations, $\langle \delta M_z^2 \rangle$, was determined from the in-situ magnetization image of the central $150 \mu$m $\times$ $150 \mu$m region of the condensate. The contribution of HQV ferromgnetic cores to $\langle \delta M_z^2 \rangle$, $\langle \delta M_z^2 \rangle_\textrm{HQV}$, was estimated by modeling the magnetization profile of a HQV core with a Gaussian function of $M_z(r)=\pm M_{z0} e^{-r^2/r_c^2}$, and $\langle \delta M_z^2\rangle_\textrm{HQV}=n_v \int M_z^2(r) 2\pi r dr=\frac{\pi}{2}n_v M_{z0}^2 r_c^2$. The peak magnetization $|M_{z0}|$ and the core radius $r_c$ were determined from a model function fit to the averaged in-situ magnetization image of the HQV core, giving $r_c=1.44\xi_s$~\cite{Seo}. We set $|M_{z0}|=1$ in our image normalization, and $\langle \delta M_z^2\rangle_{\text{HQV}}=3.26 \times n_v \xi_s^2$.

\begin{figure*}
\includegraphics[width=13cm]{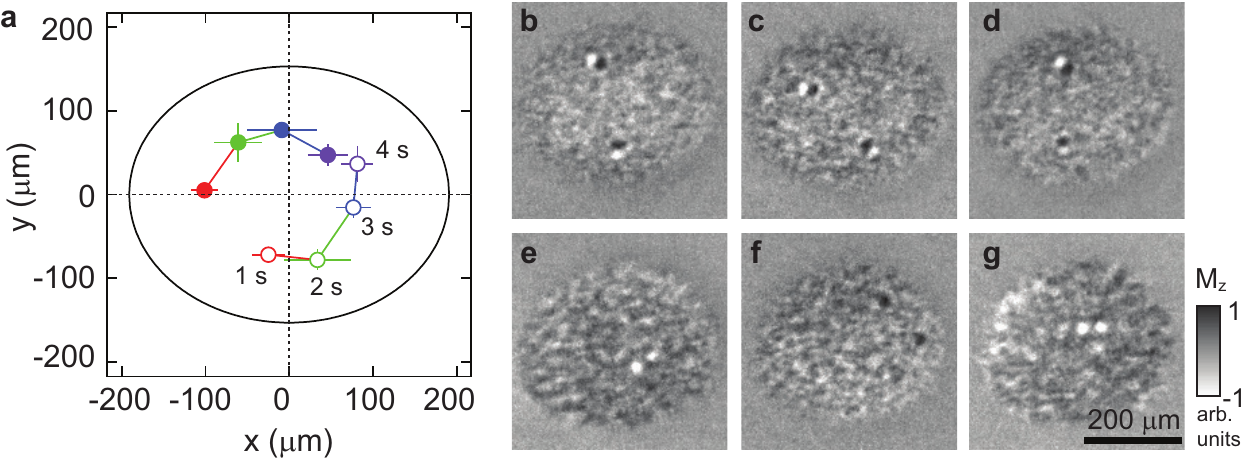}
\caption{{\bf Trajectories of the HQV pairs}. (a) Center positions of the two HQV pairs at various evolution times: $t=1$~s (red), 2~s (green), 3~s (blue), and 4~s (purple). Solid (open) circles indicate the center position of the HQV pair with clockwise (counterclockwise) mass circulation, i.e., $q_n<0$ ($q_n>0$). The center position were determined with the average from 5 data images for $t=1$~s and 10 data images for $t=2,3,4$~s, and the error bar indicates the standard deviations of the measurements from the data images. (b-c) Examples of the images for $t=2$~s. (e-g) Images, taken at $t=5$~s, showing only one HQV dipoles remaining after the HQV collision. The occurrence probability of this event was about $3\%$ in our collision experiment.}
\label{FigS2}
\end{figure*}


\begin{references}

\bibitem{RJD} R.~J.~Donnelly,~\textit{Quantized Vortices in Helium II} (Cambridge University Press, Cambridge, England, 1991).
\bibitem{Volovik} M.~M.~Salomaa, and~G.~E.~Volovik, Quantized vortices in superfluid He III,~Rev.~Mod.~Phys. {\bf59}, 533 (1987).

\bibitem{Lagoudakis} K.~G.~Lagoudakis,~T.~Ostatnicky,~A.~V.~Kavokin,~Y.~G.~Rubo,~R.~Andr\'{e}, and~B.~Deveaud-Pl\'{e}dran, Observation of half-quantum vortices in an exciton-polariton condensate,~Science {\bf326}, 974 (2009).
\bibitem{Manni} F.~Manni,~K.~G.~Lagoudakis,~T.~C.~H.~Liew,~R.~Andr\'{e}, V. Savona, and~B.~Deveaud, Dissociation dynamics of singly charged vortices into half-quantum vortex pairs,~Nat. Commum. {\bf3}, 1309 (2012).
\bibitem{Jang} J.~Jang,~D.~G.~Ferguson,~V.~Vakaryuk,~R.~Budakian,~S.~B.~Chung,~P.~M.~Goldbart, and Y.~Maeno, Observation of half-height magnetization steps in Sr$_2$RuO$_4$,~Science {\bf331}, 186 (2011).
\bibitem{Seo} S.~W.~Seo,~S.~Kang,~W.~J.~Kwon, and~Y.~Shin, Half-quantum Vortices in an Antiferromagnetic Spinor Bose-Einstein Condensate,~Phys.~Rev.~Lett. {\bf115}, 015301 (2015).


\bibitem{Autti} S.~Autti,~V.~V.~Dmitriev,~V.~B.~Eltsov,~J.~M\"{a}kinen, G. E. Volovik, A.~N.~Yudin, and V.~V.~Zavjalov, Observation of half-quantum vortices in superfluid $^{3}$He, arXiv:1508.02197.


\bibitem{HQVcoresize} A.~C.~Ji,~W.~M.~Liu,~J.~L.~Song, and F.~Zhou, Dynamical Creation of Fractionalized Vortices and Vortex Lattices,~Phys.~Rev.~Lett. {\bf101}, 010402 (2008).

\bibitem{Tsubota} H.~Takeuchi,~S.~Ishino, and M.~Tsubota, Binary Quantum Turbulence Arising from Countersuperflow Instability in Two-Component Bose-Einstein Condensates, Phys.~Rev.~Lett. {\bf105}, 205301 (2010).

\bibitem{Eto} M.~Eto,~K.~Kasamatsu,~M.~Nitta, H.~Takeuchi, and M.~Tsubota, Interaction of half-quantized vortices in two-component Bose-Einstein condensates,~Phys.~Rev.~A {\bf83}, 063603 (2011).

\bibitem{Kaneda} T.~Kaneda, and H.~Saito, Dynamics of a vortex dipole across a magnetic phase boundary in a spinor Bose-Einstein condensate,~Phys.~Rev.~A {\bf90}, 053632 (2014).

\bibitem{Kasamatsu} K.~Kasamatsu,~M.~Eto, and M.~Nitta, Short range intervortex interaction and interacting dynamics of half-quantized vortices in two-component Bose-Einstein condensates,~Phys.~Rev.~A {\bf93}, 013615 (2016).

\bibitem{Volovik2} G.~E.~Volovik, {\it The Universe in a Helium Droplet} (Oxford University Press, New York, 2003).

\bibitem{Choi} J.~Choi, W.~J.~Kwon, and Y.~Shin, Observation of Topologically Stable 2D Skyrmions in an Antiferromagnetic Spinor Bose-Einstein Condensate,~Phys.~Rev.~Lett. {\bf108}, 035301 (2012).

\bibitem{Ruostekoski} J.~Ruostekoski, and J.~R.~Anglin, Monopole Core Instability and Alice Rings in Spinor Bose-Einstein Condensates,~Phys.~Rev.~Lett. {\bf91}, 190402 (2003).
\bibitem{Ray} M.~W.~Ray, E.~Ruokokoski, K.~Tiurev, M.~M\"{o}tt\"{o}nen, and D.~S.~Hall, Observation of isolated monopoles in a quantum field, Science {\bf348}, 544 (2015).
\bibitem{Tiurev} K.~Tiurev, E.~Ruokokoski, H.~M\"{a}kel\"{a}, D.~S.~Hall, and M.~M\"{o}tt\"{o}nen, Decay of an isolated monopole into a Dirac monopole configuration, Phys.~Rev.~A {\bf93}, 033638 (2016).


\bibitem{Ho} T.-L.~Ho, Spinor Bose Condensates in Optical Traps, Phys.~Rev.~Lett. {\bf81}, 742 (1998).

\bibitem{Machida} T.~Ohmi, and K.~Machida, Bose-Einstein condensation with internal degrees of freedom in alkali atom gases, J.~Phys.~Soc.~Jpn. {\bf67}, 1822 (1998).

\bibitem{Zhou} F.~Zhou, Spin Correlation and Discrete Symmetry in Spinor Bose-Einstein Condensates,~Phys.~Rev.~Lett. {\bf87}, 080401 (2001).

\bibitem{Kawaguchi_review} Y.~Kawaguchi, and~M.~Ueda, Spinor Bose-Einstein condensates,~Phys.~Rep. {\bf520}, 253 (2012).

\bibitem{Stamper-kurn_review} D.~M.~Stamper-Kurn, and~M.~Ueda, Spinor Bose gases: Symmetries, magnetism and quantum dynamics,~Rev.~Mod.~Phys. {\bf85}, 1191 (2013).


\bibitem{Gerbier} F.~Gerbier, A.~Widera, S.~F\"{o}lling, O.~Mandel, and I.~Bloch, Resonant control of spin dynamics in ultracold quantum gases by microwave dressing,~Phys.~Rev.~A {\bf73}, 041602(R) (2006).

\bibitem{Quenching} E.~M.Bookjans,~A.~Vinit, and C.~Raman, Quantum Phase Transition in an Antiferromagnetic Spinor Bose-Einstein Condensate,~Phys.~Rev.~Lett. {\bf107}, 195306 (2011).

\bibitem{Zhao}  L.~Zhao, J.~Jiang, T.~Tang, M.~Webb, and Y.~Liu, Dynamics in spinor condensates tuned by a microwave dressing field,~Phys.~Rev.~A {\bf89}, 023608 (2014).


\bibitem{Supple} See Supplemental Material for experimental details on the sample preparation, the vortex state preparation, the magnetization imaging, and the analysis of $\langle \delta M_z^2\rangle_{\text{HQV}}$. Additional data on the HQV-pair collision and the vortex decay for the EAP phase and Ref.~\cite{Heo} are included.

\bibitem{Heo} M.-S.~Heo, J. Choi, and Y.~Shin, Fast production of large $^{23}$Na Bose-Eintstein condensates in an optically plugged magnetic quadrupole trap,~Phys.~Rev.~A {\bf 83}, 013622 (2011).

\bibitem{Higbie} J.~M.~Higbie, L.~E.~Sadler, S.~Inouye, A.~P.~Chikkatur, S.~R.~Leslie, K.~L.~Moore, V.~Savalli, and D.~M.~Stamper-Kurn, Direct Nondestructive imaging of Magnetization in a spin-1 Bose-Einstein Gas,~Phys.~Rev.~Lett. {\bf95}, 050401 (2005).

\bibitem{Neely} T.~W.~Neely, E.~C.~Samson, A.~S.~Bradley, M.~J.~Davis, and B.~P.~Anderson, Observation of Vortex Dipoles in an Oblate Bose-Einstein Condensate,~Phys.~Rev.~Lett. {\bf104}, 160401 (2010).

\bibitem{Freilich} D.~V.~Freilich, D.~M.~Bianchi, A.~M.~Kaufman, T.~K.~Langin, and D.~S.~Hall, Real-time dynamics of single vortex lines and vortex dipoles in a Bose-Einstein condensate, Science {\bf329}, 1182 (2010).

\bibitem{Kwon1} W.~J.~Kwon,~S.~W.~Seo, and Y.~Shin, Periodic shedding of vortex dipoles from a moving  penetrable obstacle in a Bose-Einstein condensate,~Phys.~Rev.~A {\bf92}, 033613 (2015).

\bibitem{Kwon2} W.~J.~Kwon, G.~Moon, J.~Choi, S.~W.~Seo, and Y.~Shin, Relaxation of superfluid turbulence in highly oblate Bose-Einstein condensates,~Phys.~Rev.~A {\bf90}, 063627 (2014).
\bibitem{Symes} L.~M.~Symes,~D.~Baillie, and P.~B.~Blakie, Static structure factors for a spin-1 Bose-Einstein condensate,~Phys.~Rev.~A {\bf 89}, 053628 (2014).


\bibitem{Mermin66} N.~D.~Mermin and H.~Wagner, Absence of Ferromagnetism or Antiferromagnetism in One- or Two-dimensional Isotropic Heisenberg Models,~Phys.~Rev.~Lett. {\bf17}, 1133 (1966).
\bibitem{Nelson1977} D.~R.~Nelson and R.~A.~Pelcovits, Momentum-shell recursion relations, anisotropic spins, and liquid crystals in 2+$\epsilon$ dimensions,~Phys.~Rev.~B {\bf16}, 2191 (1977).

\bibitem{Mukerjee} S.~Mukerjee,~C.~Xu, and J.~E.~Moore, Topological Defects and the Superfluid Transition of the s=1 Spinor Condensate in Two Dimensions,~Phys.~Rev.~Lett. {\bf97}, 120406 (2006).
\bibitem{Pietila} V.~Pietil\"a,~T.~P.~Simula, and M.~M\"ott\"onen, Finite-temperature phase transitions in quasi-two-dimensional spin-1 Bose gases,~Phys.~Rev.~A {\bf81}, 033616 (2010).
\bibitem{Phase_Diagram} A.~J.~A.~James and A.~Lamacraft, Phase Diagram of Two-Dimensional Polar Condensates in a Magnetic Field,~Phys.~Rev.~Lett. {\bf106}, 140402 (2011).

\end{references}
\end{document}